\documentclass[10pt,sigconf,nonacm]{acmart}
\AtBeginDocument{%
  }

\setcopyright{acmlicensed}
\copyrightyear{2025}
\acmYear{2025}
\acmDOI{XXXXXXX.XXXXXXX}
\acmConference[ANRW '25]{ACM/IRTF Applied Networking Research Workshop 2025}{July 19-25, 2025}{Madrid, Spain}
\acmISBN{978-1-4503-XXXX-X/2018/06}

\usepackage{graphicx}

\begin{document}

\title{QUIC Delay Control: an implementation of congestion and delay control}

\author{Saverio Mascolo, Andrea Vittorio Balillo, Gioacchino Manfredi, \\ Davide D'Agostino, Luca De Cicco}

    \affiliation{%
  \institution{Politecnico di Bari, Dipartimento di Ingegneria Elettrica e dell'Informazione}
  \city{Bari}
  \country{Italy}}
\email{name.surname@poliba.it}

\renewcommand{\shortauthors}{Mascolo et al.}

\begin{abstract}

A new congestion and delay control algorithm named QUIC Delay Control (QUIC-DC) is proposed for controlling not only congestion but also the queueing delay encountered along the forward communication path. The core idea is to estimate the one-way queueing delay of a connection to trigger an early reaction to congestion. This idea, along with a  the   TCP Westwood+ congestion control algorithm, has been implemented in  QUIC-DC and compared  with QUIC Cubic, BBRv2, NewReno, Westwood+.
The results obtained in the emulated and real network environments show that QUIC-DC can significantly reduce packet losses along with end-to-end communication delays, while preserving network utilization, features that are both very useful for real-time applications.
\vspace{-1mm}
\end{abstract}

\ccsdesc[500]{Network protocols ~low-delay congestion control}


\keywords{Congestion Control, Delay Control, QUIC }

\maketitle

\section{Introduction }

Classic congestion control aims at maximizing throughput while minimizing congestion, i.e. packet losses and retransmissions. This implies that communication delays are not controlled and oscillate between a minimum, equal to the propagation time, up to a maximum equal to the propagation time plus the maximum queuing delay encountered along the communication path.

The goal of this paper is to propose an algorithm that not only avoids congestion while still providing high network utilization but also controls the network queueing delay. 

Controlling delays is nowadays of the utmost importance since the Internet is not only aimed at delay insensitive data traffic, but is increasingly used for  real-time applications such as live streaming, video-conferencing, virtual reality/augmented reality, $360^\circ$ video,  autonomous driving, tele-robotics and tele-surgery.

In this paper we propose a new congestion and delay control and we experiment it in controlled and uncontrolled network environments.

\begin{figure*}[t]
\begin{centering}
\includegraphics[width=\textwidth]{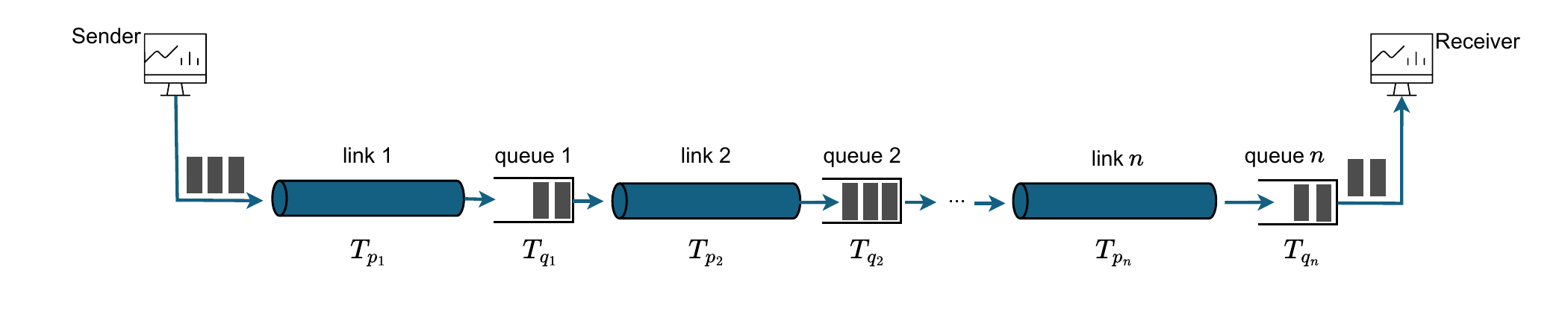}
\par\end{centering}
\vspace{-3mm}
\caption{Packets going from the sender to the receiver through a series of links and buffers}
\label{fig:framework}
\vspace{-3mm}
\end{figure*}

Figure~\ref{fig:framework} shows an Internet  connection between two TCP/UDP peers over an IP network. The essential elements that play a key role   are the communication links and the buffers. Indeed, TCP/IP or UDP/IP packets go through a set of communication links and buffers connecting the routers that are encountered as they travel  from the sender to the destination. Figure~\ref{fig:framework} shows the sequence of communication links and buffers, each link $i$ with a propagation delay $T_{p_i}$ and  each buffer $i$ with a queueing delay $T_{q_i}$, which are encountered by a packet travelling from the sender to the destination.

The round trip time ($RTT$) is the sum of the  propagation  and queueing delays encountered going from the sender to the destination and then back to the sender:  $RTT= \sum_{i\in P} (T_{p_i}+ T_{q_i})$, where $P$ is the set of links and buffers belonging to the bidirectional communication path.

The propagation delays $T_{p_i}$ are fixed and depend on the physical distance (the propagation speed is roughly 1/3 of the speed of light). In contrast, the queueing delays $T_{q_i}$ depend on queueing occupancy and can be reduced by controlling the queue length. The overall queueing delay is $T_{q}= \sum_{i\in P} T_{q_i}$, whereas the overall propagation  delay is $T_{p}= \sum_{i\in P} T_{p_i}$.

In other terms, the RTT is made of a constant propagation time plus a time-varying stochastic component due to the queueing delay as follows: $RTT= T_{p}+ T_{q}$

QUIC-DC can bound the component of delay due to the queueing. Conceptually, it can be viewed as a form  of  Explicit Congestion Notification (ECN)~\cite{ramakrishnan2001addition} implemented end-to-end.

\vspace{-3mm}
\section{Related Work}

Traditional loss-based TCP~\cite{jac88, peterson2007computer} is not suitable for delay-sensitive traffic, such as in the case of video conferencing traffic,  because its congestion control mechanism continuously probes for available network bandwidth, creating a periodic pattern in which network queues are cyclically filled up to packet loss and then drained. These queue oscillations introduce a time-varying stochastic queueing delay component, which adds to the fixed network propagation delay, thereby impairing delay-sensitive communications. But more than anything, it is the retransmission of lost packets that harms real-time communication. Therefore, reducing the packet loss rate is fundamental in real-time traffic.
For this reason, the idea of using delay measurements to proactively react to the onset of network congestion has been proposed~\cite{jain89}. Issues related to delay measurements and throughput degradation when loss-based flows share the bottleneck with delay-based flows have been addressed in~\cite{gri04,bun11,dr04}.
The first efforts focusing on the reduction of the queueing delay were set in the TCP congestion control research domain. In particular, the first congestion control algorithm specifically designed to contain the end-to-end latency by employing delay measurements is described in the seminal work by Jain~\cite{jain89}. Since then, many delay-based TCP congestion control variants have been designed, such as TCP Vegas~\cite{vegas} or ~\cite{bun11}. 
TCP Vegas employs end-to-end RTT measurements to establish delay thresholds used to infer incipient congestion by comparison with a base RTT~\cite{vegas}. It has been shown that, when the RTT is used as a congestion metric, a low channel utilization may be obtained in the presence of reverse traffic, which inflates queues in the backward path, or when competing with loss-based flows~\cite{gri04}. To tackle such fairness issues, the algorithms proposed in~\cite{bun11} infer the presence of concurrent loss-based flows and switch to loss-based mode, returning back to the delay-based mode when loss-based flows are detected to be no longer present.
Other algorithms employ one-way-delay measurements to rule out the sensitivity to reverse-path congestion. Examples are LEDBAT (over UDP)~\cite{sha12} and TCP Santa Cruz~\cite{parsa1999improving}. In particular, LEDBAT employs an increasing phase in which the congestion window is increased at a rate that is proportional to the distance between the measured one-way delay and a fixed delay target~\cite{sha12}. It has been shown that LEDBAT is affected by the so-called \textit{latecomer effect}: when two flows share the same bottleneck the second flow typically starves the first one~\cite{car10}.
Recently the idea of employing RTT gradient has been proposed to overcome the aforementioned latecomer effect. Some examples are CDG~\cite{hay11} and Verus~\cite{zak15}. CDG has been designed with the aim of coexisting with loss-based flows while keeping end-to-end delay low~\cite{hay11}. 
Verus has been designed for the case of cellular networks where sudden link capacity variations make the congestion control design challenging \cite{zak15}. 
The Google Congestion Control (GCC)~\cite{gcc},  implemented in the Google Chrome web browser, employs the one way delay variation as a congestion signal and an adaptive threshold mechanism to effectively compete with loss-based congestion control algorithms~\cite{gcc}. GCC is the only widely deployed non-proprietary congestion control algorithm used in a video conferencing application such as Google  Meet.

\section{QUIC Delay Control}

In this section, we introduce the main idea to obtain queueing delay control.  The algorithm  will be implemented and evaluated in QUIC~\cite{rfc9000}.


To focus on the  delay problem, let us consider again  Figure~\ref{fig:framework}, which shows a connection between two TCP/UDP peers over an IP network. As we have seen, the RTT is the sum of the encountered propagation delays and queueing delays.

%

Therefore, by  keeping  the queue as low as possible, we can target a minimum RTT equal to the propagation delay only:  
 \begin{equation}
     \ {RTT}_{\text{min}} \ = T_{p}
 \end{equation}

\noindent Similarly, $\text{RTT}_{\text{max}}$ can be defined as:
 \begin{equation}
      {RTT}_{\text{max}}   = {RTT}_{\text{min}}+ T_{q_{\max} }
 \end{equation}
\noindent where $T_{q_{\max}}$ is the maximum queueing time encountered by a packet going through all the buffers belonging to the communication path.
%

\begin{figure}
\begin{centering}
\includegraphics[width=0.45\textwidth]{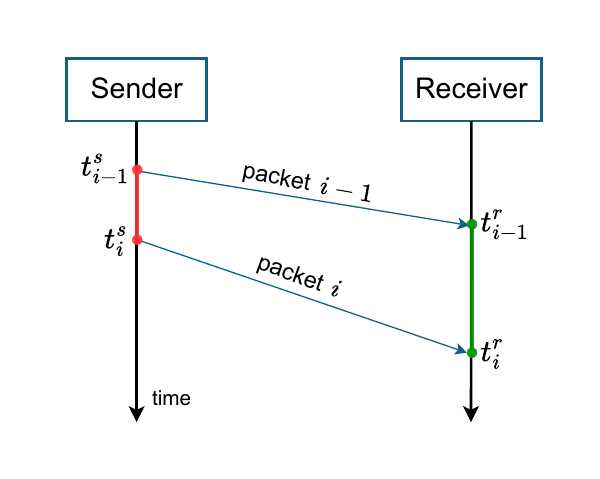}
\par\end{centering}
\vspace{-3mm}
\caption{Measurement of one-way delay variation}
\label{fig:owd}
\vspace{-3mm}
\end{figure}

At this point, it is important to distinguish between queueing on the forward path and queueing on the backward path. A connection is primarily constrained by congestion on the forward path; therefore, if RTT is used as a congestion signal, queueing in the backward path caused by reverse traffic may lead to a false indication of congestion. This, in turn, would unnecessarily slow down the forward traffic — a behavior that is indeed observed in TCP Vegas~\cite{gri04}.

To this end, it is necessary to consider the variation of the One-Way Delay (OWD), where the OWD is  the time a packet takes to travel from the sender to the receiver,  in order to estimate the forward queueing delay only.
Precisely, the OWD  is the difference between the packet arrival time $t^{r}$ at the receiver and the packet departure time $t^{s}$ at the sender as shown in Figure \ref{fig:owd}:
\begin{equation}
OWD_{i} = t^r_i- t^s_i
\end{equation}
The OWD measurement contains a systematic error due to the clock offset between the sender and the receiver. 
However,  what is useful to know is not the absolute value but just the OWD variation $OWDV$, which measures an increase or decrease  in the \textit{forward queueing delay}.
Thus, by computing the \textit{one-way delay variation}:
\begin{equation}
\begin{split}
 & OWDV_{i} ={OWD_{i}-OWD_{i-1}} \\
 &= (t^r_i- t^s_i) - (t^r_{i-1} - t^s_{i-1}) \\
 &= (t^r_i-t^r_{i-1}  ) - (t^s_i - t^s_{i-1})
\end{split}
\end{equation}

Equation 4 shows that the \textit{one-way delay variation} can be easily obtained by computing the difference between the inter arrival times of two consecutive packets $i$ and $i$–$1$, and their corresponding inter departure times.

By summing the one way delay variations, we can compute the  one-way queueing delay as:
\begin{equation}
OWQD_{i}= OWQD_{i-1} + OWDV_i
\end{equation}

It should be noted that this measure is both  simple and robust and can be easily implemented. 

Once we have  the  measurement   of the  queueing delay along the forward path, we can use it to define \textit{a new congestion event}  which occurs when $OWQD$ reaches a threshold $OWQD_{th}$:
\begin{equation}
OWQD > OWQD_{th}
\label{eq:owdth}
\end{equation}
%
Using Eq. (\ref{eq:owdth}), we can control queue build-up and reduce the risk of packet loss, which is particularly important for latency-sensitive applications.

In QUIC Delay Control (QUIC-DC), we use the TCP Westwood+ congestion control~\cite{gri04,mascolo2001tcp} i.e., after a congestion episode,  QUIC-DC  sets  the  congestion window equal to the product of the available bandwidth times the minimum $RTT$, which   keeps full the propagation pipe and empty the queueing pipe.

In summary, a congestion episode is defined by a timeout, a 3DUPACK, or $\text{OWD}>\text{OWD}_{th}$. 
The reaction to a congestion episode is the same of that of TCP Westwood+ after a timeout or after  3DUPACK, and the reaction to  $\text{OWD}>\text{OWD}_{th}$ is the same of TCP Westwood+ after 3DUPACK. 

The implementation of TCP Westwood+ congestion control~\cite{gri04} in QUIC-DC closely adheres to its  counterpart  implementation in the Linux TCP kernel with a small change in  the Westwood+ low-pass filter to make the bandwidth estimation BWE faster. Indeed, the original  low pass filter  used to estimate the available bandwidth BWE adversely impacts the algorithm's responsiveness during transient network conditions, leading to a lower goodput during the start up phase. 
The low pass filter employed in QUIC-DC is:
\begin{equation}
BWE_{i}= 0.2 \cdot BWE_{i-1} + 0.8 \cdot bandwidthsample_{i}
\end{equation}
\noindent which significantly improves the goodput. 



 

\section{QUIC-DC implementation and the testbed}
\label{sec:testbed}

QUIC-DC has been implemented into Meta's \texttt{mvfst} QUIC transport implementation, which serves as the underlying transport layer for \texttt{Proxygen}, Meta's HTTP libraries collection\footnote{\url{https://engineering.fb.com/2014/11/05/production-engineering/introducing-proxygen-facebook-s-c-http-framework}}. Both \texttt{mvfst} and \texttt{Proxygen} are implemented in C++ to ensure high performance and scalability. To simplify deployment and testing, all components were containerized using Docker.
In our laboratory testbed, we have set the maximum possible value for the advertised window so that the congestion window size is constrained only by the congestion window. 

 QUIC-DC has been evaluated using two testbeds. The first is a controlled laboratory testbed made of two machines connected with an Ethernet cable. The first machine acts as the client, sending one or multiple HTTP3/QUIC requests to the second machine, which acts as the server. Both machines are equipped with Gigabit Ethernet Network Interface Controllers (NICs). All the Docker containers are based on Ubuntu 22.04 images. The bottleneck is located physically on the interface of the server where we applied a 25\,ms delay both incoming and outgoing, thus summing up to a total propagation time  of 50\,ms. We use the Linux traffic control (\texttt{tc}) tool from the \texttt{iproute2} suite to perform all the network emulation tasks\footnote{\url{https://man7.org/linux/man-pages/man8/tc.8.html}}. The propagation delay $T_p$ is set using \texttt{netem} queueing disciplines on the server Ethernet interface, while the bandwidth capacity $C$ in Mbps and the buffer size $B$ in bytes are set using the token bucket filter (TBF) mechanism. The results have been averaged over 3 runs and appear consistent between each run.


\begin{figure}
\begin{centering}
\includegraphics[width=0.4\textwidth]{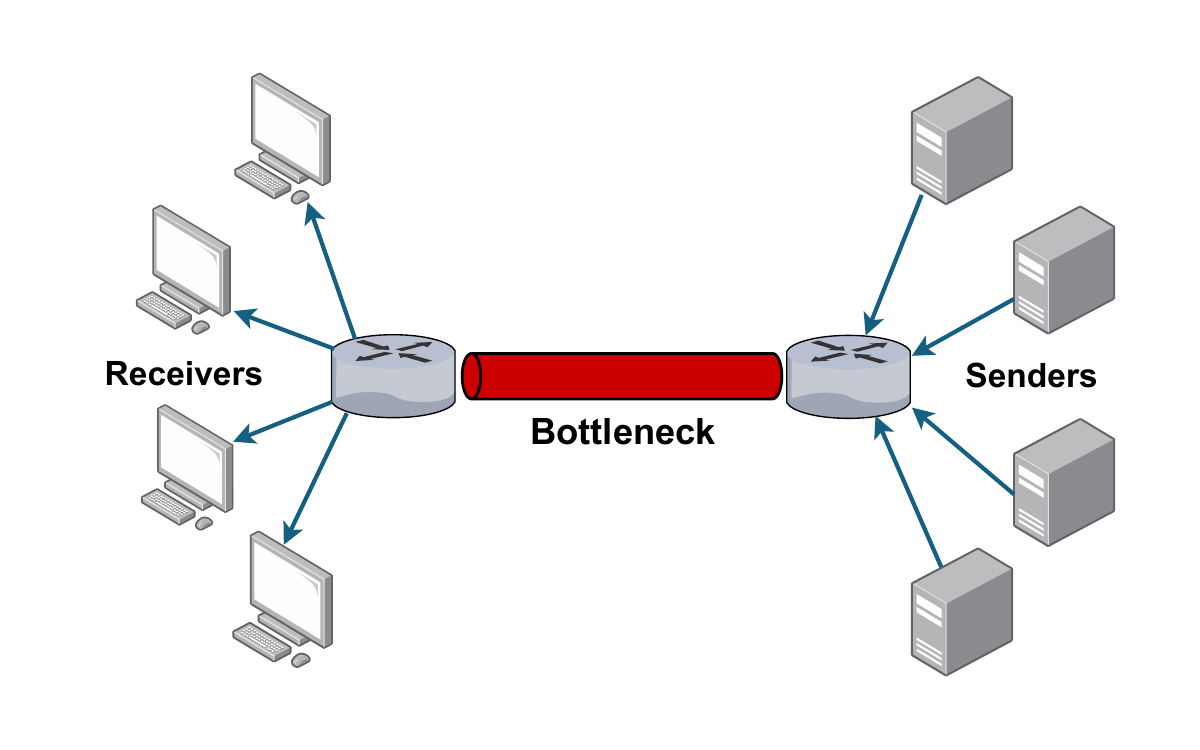}
\par\end{centering}
\vspace{-3mm}
\caption{Single bottleneck testbed}
\label{fig:bottle}
\vspace{-3mm}
\end{figure}

The other testbed considered is an uncontrolled real Internet setup consisting of a residential client in Bari that accesses the Internet using an ADSL modem. The server is located at the University of Rome, La Sapienza.

Pacing is enabled at the application level to ensure that packets are transmitted at evenly spaced intervals, minimizing burstiness. \footnote{The pacing interval has been set to 200 $\mu s$.}

\section{Experimental results}
QUIC-DC has been  compared with other state-of-the-art congestion control algorithms, namely Westwood+~\cite{ mascolo2001tcp, gri04}, Cubic~\cite{ha2008cubic}, New Reno~\cite{henderson2012newreno}, and BBR version 2 ~\cite{cardwell2019bbr}.In this work   BBRv2 has been considered due to its superior fairness compared to BBRv3~\cite{zey24}.
The bottleneck buffer size $B$ has been set equal {0.5, 1, 2, 4} times the bandwidth delay product  (BDP),  where the bandwidth is the bottleneck capacity $C$ and the delay is the round trip propagation delay $T_{p}=RTT_{min}$. We will consider  single-flow and multiple-flows scenarios.

\subsection{Emulated network results}
\label{subsec:emulresults}
\subsubsection{Single Flow}
To understand the dynamics of QUIC-DC, we begin by examining a single-flow connection transferring a 100\,MB file.  QUIC-DC has been tested using different values for the one-way queueing delay threshold   $OWQD_{th}$, which has been set equal to 10\%, 20\%, 50\%, and 80\% of the bottleneck buffer size expressed in time.


\begin{figure}
\begin{centering}
\includegraphics[width=0.45\textwidth]{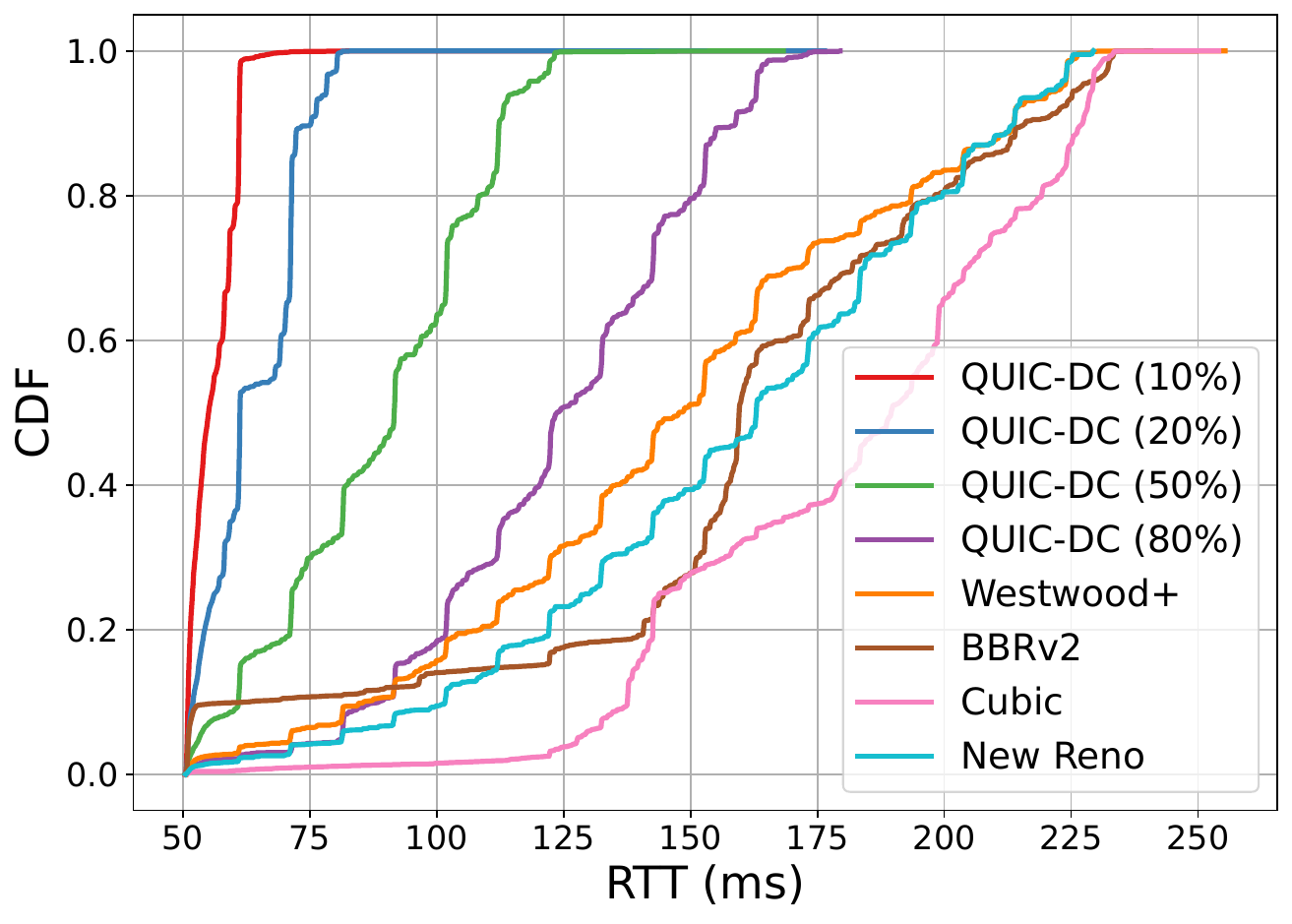}
\par\end{centering}
\vspace{-3mm}
\caption{Round‑trip‑time (RTT) distribution for a single flow – emulated link, buffer=4×BDP}
\label{fig:cdf}
\vspace{-4mm}
\end{figure}

Table~\ref{tab:4BDP} shows the performance metrics of a single flow connections for the case of a buffer size set equal to 4 BDP, where $C$=25\,Mbps and $RTT_{min}=T_{p}=$50\,ms. In this case, all the congestion control algorithms achieve comprable goodputs but with larger loss rate in the case of Cubic and BBRv2.  In comparison to QUIC Cubic, the loss percentage is reduced up to two orders of magnitude. 

Figure~\ref{fig:cdf} shows the corresponding CDF of the measured RTT for all the algorithms under analysis. The curves show that QUIC-DC provides lower RTT values compared to the other algorithms. Moreover, it can be noticed that by decreasing the $OWQD_{th}$ the  RTT curves shift to the left towards lower RTT values. This is expected since a higher $OWQD_{th}$  implies larger queues along the path. It is worth noting that in the case of algorithms with a larger loss rate, the RTT measurements underestimate the effective delays because retransmitted packets experience very large delays, which do not contribute to the RTT measurements \cite{karn1987improving}.
Similar considerations can be drawn from Table~\ref{tab:2BDP}, which corresponds to the case of bottleneck buffer size  set to 2 BDP.

\begin{figure} 
\begin{centering}
\includegraphics[width=0.45\textwidth]{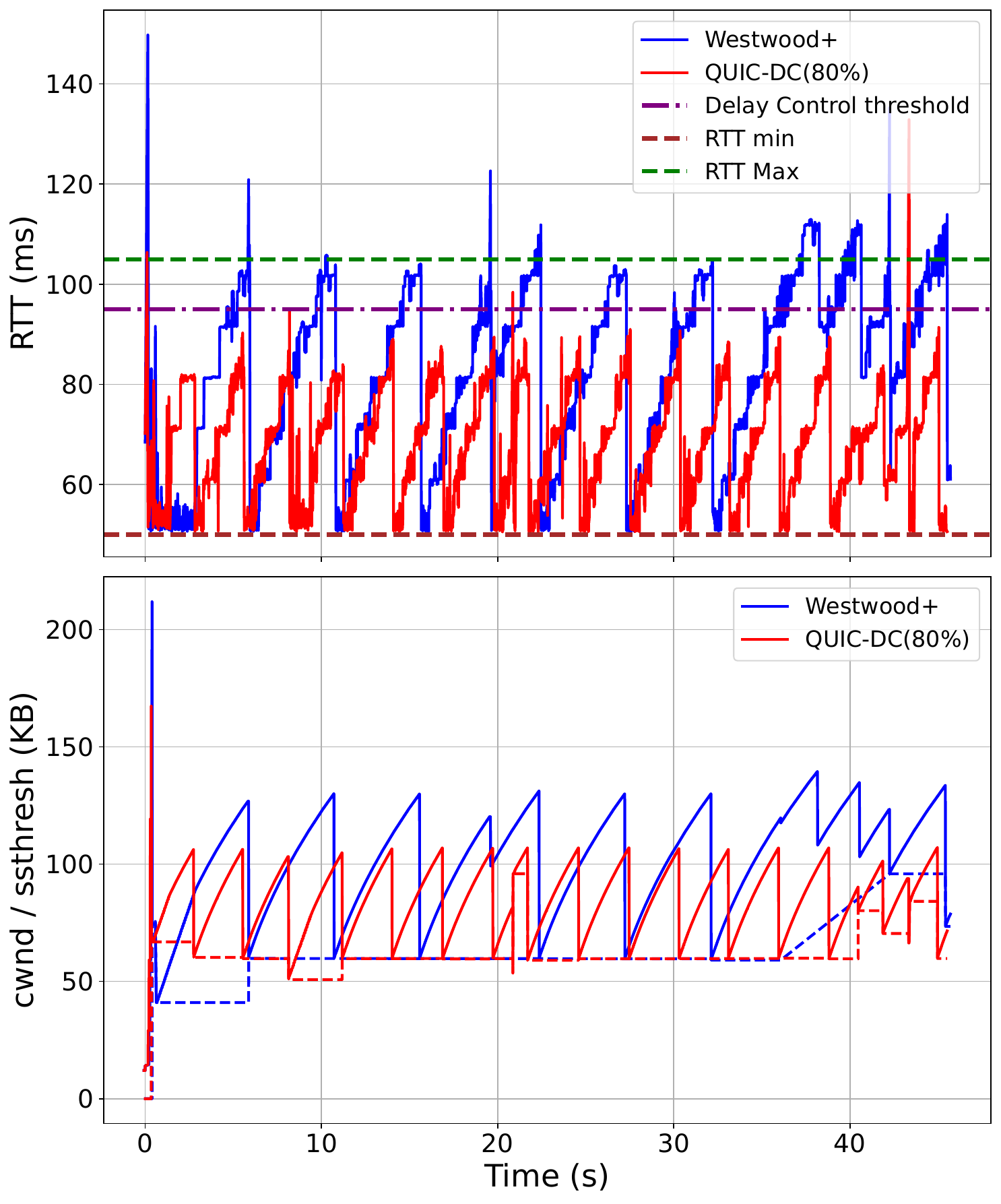}
\par\end{centering}
\vspace{-1mm}
\caption{Time evolution of RTT and congestion window  – buffer=2×BDP. }
\label{fig:saw}
\end{figure}

A nice understanding of the meaning of the proposed approach is provided by  Figure~\ref{fig:saw}, which compares the  $RTT$ and  $cwnd$ dynamics of QUIC Westwood+ (blue line) and QUIC-DC (red line) in the case of a bottleneck buffer size set to 2 BDP. It can be observed that, while QUIC Westwood+ resets the congestion window when the maximum RTT is reached (dashed green line), QUIC-DC resets the congestion window when the $OWQD$ reaches the delay control threshold set at 80\% of the bottleneck buffer size in time. Under this setting, buffer saturation at the bottleneck is prevented, effectively avoiding packet losses. Consequently, the observed packet loss ratio was reduced by one order of magnitude (see Table~\ref{tab:4BDP}).


When the buffer size is set to 1 BDP or 0.5 BDP, the results in Table \ref{tab:1BDP} and Table \ref{tab:0.5BDP} indicate that QUIC-DC tends to be less efficient in terms of goodput. Notably, in this scenario with an undersized bottleneck buffer, both BBRv2 and Cubic achieve higher goodput, with Cubic also exhibiting the lowest loss rate. As will be shown in the following, this result disappears in the case of a multi-flow scenario.
\vspace{-3mm}

\begin{table}
  \centering
  \caption{Single-flow emulation; buffer = $4\times\text{BDP}$
           ( $C=10\,\text{Mbps}$, $RTT_{\min}=50\,\text{ms}$ ).}
  \label{tab:4BDP}

  \resizebox{0.48\textwidth}{!}{%
    \begin{tabular}{lrrrrr}
      \toprule
      \textbf{CCA} & \textbf{Gput} & \textbf{Tput} & \textbf{Loss}
                   & \textbf{RTT\textsubscript{avg}}
                   & \textbf{RTT\textsubscript{std}}\\
      & (Mbps) & (Mbps) & (\%) & (ms) & (ms)\\
      \midrule
QUIC-DC (10\%) & 9.16 & 9.16 & 0.02 & \textbf{55.75} & 3.00 \\
QUIC-DC (20\%) & 9.17 & 9.17 & 0.01 & 63.83 & 7.64 \\
QUIC-DC (50\%) & 9.16 & 9.17 & 0.03 & 88.60 & 19.14 \\
QUIC-DC (80\%) & \textbf{9.16} & 9.16 & \textbf{0.02} & 123.82 & 26.48 \\
Westwood+       & 9.18 & 9.19 & 0.07 & 170.23 & 43.10 \\
BBRv2           & 9.15 & 9.21 & 0.65 & 102.33 & 27.44 \\
Cubic           & 9.10 & 9.20 & 1.03 & 190.32 & 31.37 \\
New Reno        & 9.20 & 9.20 & 0.07 & 159.26 & 45.14 \\

\bottomrule
\end{tabular}
}
\end{table}

\begin{table}
  \centering
  \caption{Single-flow emulation; buffer = $2\times\text{BDP}$
           ( $C=10\,\text{Mbps}$, $RTT_{\min}=50\,\text{ms}$ ).}
  \label{tab:2BDP}

  \resizebox{0.48\textwidth}{!}{%
    \begin{tabular}{lrrrrr}
      \toprule
      \textbf{CCA} & \textbf{Gput} & \textbf{Tput} & \textbf{Loss}
                   & \textbf{RTT\textsubscript{avg}}
                   & \textbf{RTT\textsubscript{std}}\\
      & (Mbps) & (Mbps) & (\%) & (ms) & (ms)\\
      \midrule
QUIC-DC (10\%) & 9.06 & 9.06 & 0.03 & \textbf{52.74} & 1.23 \\
QUIC-DC (20\%) & 9.16 & 9.16 & 0.03 & 55.98 & 3.65 \\
QUIC-DC (50\%) & 9.14 & 9.15 & 0.14 & 68.33 & 10.25 \\
QUIC-DC (80\%) & \textbf{9.16} & 9.16 & \textbf{0.04} & 84.86 & 16.47 \\
Westwood+       & 9.15 & 9.18 & 0.26 & 90.53 & 20.04 \\
BBRv2           & 9.12 & 9.18 & 0.66 & 87.94 & 26.94 \\
Cubic           & 9.16 & 9.20 & 0.39 & 104.82 & 13.57 \\
New Reno        & 9.17 & 9.18 & 0.14 & 93.78 & 18.79 \\
\bottomrule
\end{tabular}
}
\end{table}

\begin{table}
  \centering
  \caption{Single-flow emulation; buffer = $1\times\text{BDP}$
           ( $C=10\,\text{Mbps}$, $RTT_{\min}=50\,\text{ms}$ ).}
  \label{tab:1BDP}

  \resizebox{0.48\textwidth}{!}{%
    \begin{tabular}{lrrrrr}
      \toprule
      \textbf{CCA} & \textbf{Gput} & \textbf{Tput} & \textbf{Loss}
                   & \textbf{RTT\textsubscript{avg}}
                   & \textbf{RTT\textsubscript{std}}\\
      & (Mbps) & (Mbps) & (\%) & (ms) & (ms)\\
      \midrule
QUIC-DC (10\%) & 7.08 & 7.13 & 0.63 & 52.51 & 1.06 \\
QUIC-DC (20\%) & 7.75 & 7.79 & 0.53 & 52.63 & 1.65 \\
QUIC-DC (50\%) & 8.51 & 8.58 & 0.83 & 53.64 & 3.96 \\
QUIC-DC (80\%) & 8.39 & 8.46 & 0.83 & 52.44 & 1.15 \\
Westwood+       & 8.22 & 8.27 & 0.64 & 53.11 & 3.03 \\
BBRv2           & 9.12 & 9.21 & 0.87 & 57.21 & 7.72 \\
Cubic           & 9.19 & 9.20 & 0.15 & 63.06 & 7.15 \\
New Reno        & 7.39 & 7.42 & 0.37 & 52.39 & 1.10 \\
\bottomrule
\end{tabular}
}
\vspace{-2mm}
\end{table}

\begin{table}
  \centering
  \caption{Single-flow emulated link; buffer = $0.5\times\text{BDP}$
           ( $C=10\,\text{Mbps}$, $RTT_{\min}=50\,\text{ms}$ ).}
  \label{tab:0.5BDP}

  \resizebox{0.48\textwidth}{!}{%
    \begin{tabular}{lrrrrr}
      \toprule
      \textbf{CCA} & \textbf{Gput} & \textbf{Tput} & \textbf{Loss}
                   & \textbf{RTT\textsubscript{avg}}
                   & \textbf{RTT\textsubscript{std}}\\
      & (Mbps) & (Mbps) & (\%) & (ms) & (ms)\\
      \midrule
QUIC-DC (10\%) & 3.30 & 3.32 & 0.44 & 52.45 & 1.49 \\
QUIC-DC (20\%) & 3.58 & 3.62 & 1.24 & 52.23 & 1.06 \\
QUIC-DC (50\%) & 3.44 & 3.49 & 1.42 & 52.11 & 1.02 \\
QUIC-DC (80\%) & 3.75 & 3.82 & 1.77 & 52.26 & 1.04 \\
Westwood+       & 3.68 & 3.73 & 1.33 & 52.29 & 1.02 \\
BBRv2           & 6.27 & 6.35 & 1.20 & 51.41 & 0.77 \\
Cubic           & 5.53 & 5.54 & 0.24 & 51.32 & 0.95 \\
New Reno        & 3.34 & 3.37 & 0.81 & 52.48 & 0.96 \\
\bottomrule
\end{tabular}
}
\vspace{-2mm}
\end{table}

\subsubsection{Multiple Flows}

The  multi-flow scenario  shown in Figure~\ref{fig:bottle} is considered. In particular, four connections of the same type  are started at different  times to avoid synchronization. In this case, the control delay threshold $OWQD_{th}=80\%$ and the bottleneck buffer size is set to 2 BDP. We have measured and averaged goodputs, throughputs, losses and RTTs for each connection. 

\begin{table}
\centering
\caption{Four concurrent flows, emulated link – buffer=2×BDP. (Average $\pm$ standard deviation).}
\label{tab:aggMetrics}
\resizebox{0.48\textwidth}{!}{
\begin{tabular}{lccc}
\toprule
\textbf{CCA}         & \textbf{RTT$_{avg}$} & \textbf{Goodput}  & \textbf{Loss} \\
                     & (ms)                 &  (Mbps, JFI)      & (\%) \\
\midrule
QUIC-DC (80\%)   & $\boldsymbol{95.07} \pm \boldsymbol{0.16}$   & $2.29 \pm 0.07$ (0.999)   & $\boldsymbol{0.12} \pm \boldsymbol{0.02}$ \\
Westwood+        & $101.19 \pm 0.13$  & $2.28 \pm 0.12$ (0.998)   & $0.43 \pm 0.06$ \\
BBRv2            & $111.55 \pm 0.29$  & $2.28 \pm 0.34$ (0.984)   & $1.05 \pm 0.11$ \\
Cubic            & $114.26 \pm 0.09$  & $2.30 \pm 0.03$ (0.999)   & $0.34 \pm 0.03$ \\
New Reno         & $113.87 \pm 0.44$  & $1.95 \pm 0.12$ (0.998)   & $0.58 \pm 0.07$ \\
\bottomrule
\end{tabular}
}
\vspace{-2mm}
\end{table}

Table \ref{tab:aggMetrics} presents the average values along with the corresponding standard deviations for the RTT, goodput, Jain's fairness index, and loss percentage of each control algorithm. As shown, QUIC-DC exhibits the lowest RTT and loss percentage, while maintaining a goodput level comparable to the other considered algorithms. It is worth noting that, since the link capacity is set to 25 Mbps, a goodput  around 6\,Mbps indicates that the link is fully utilized and fairly shared among the four connectioncs.

\begin{figure}
\begin{centering}
\includegraphics[width=0.42\textwidth]{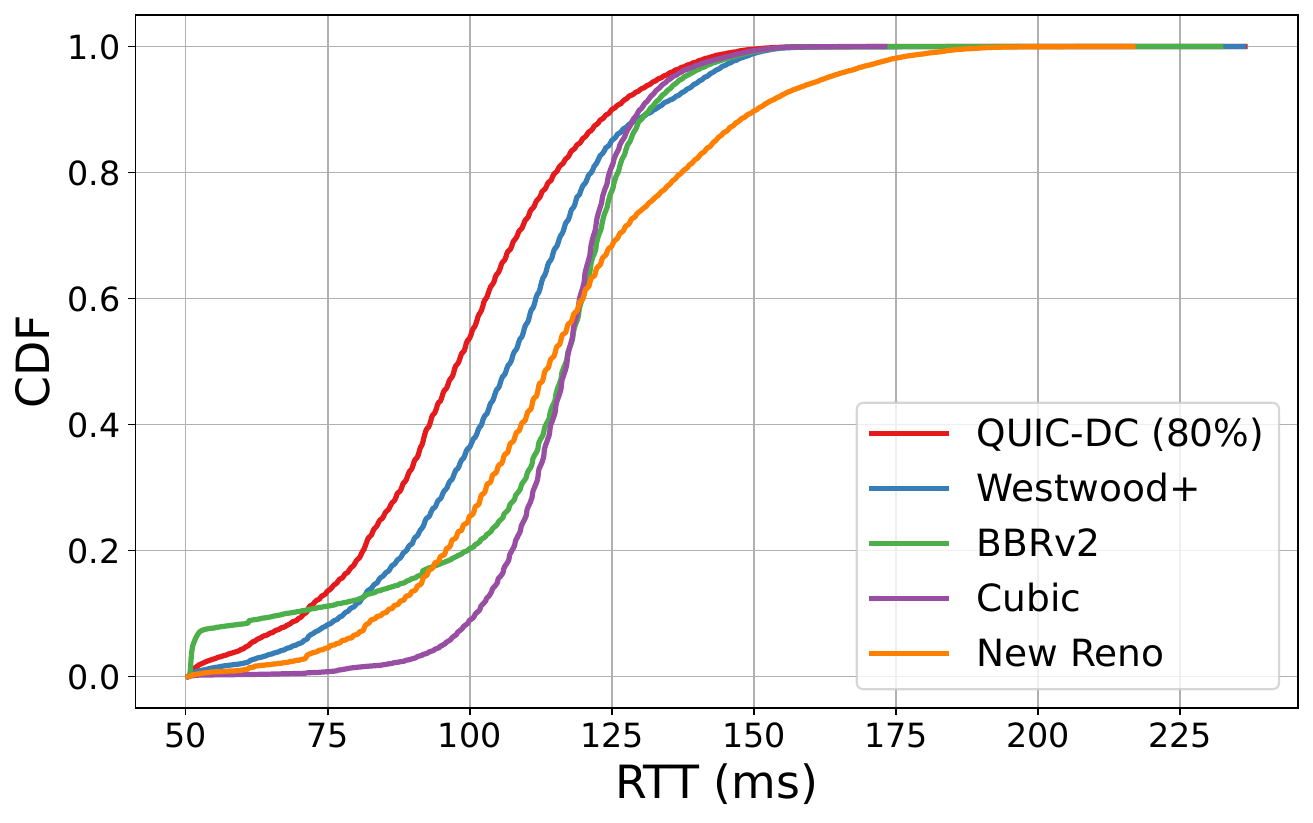}
\par\end{centering}
\vspace{-3mm}
\caption{RTT distribution for four concurrent flows – emulated link, buffer=2×BDP. }
\label{fig:bottlemulti}
\vspace{-4mm}
\end{figure}

Figure~\ref{fig:bottlemulti} shows the  CDF of the RTT in the case of multiple connections over a shared bottleneck. It shows that QUIC-DC provides lower   RTT values compared to the other considered congestion control algorithms.
Due to space constraints, we are here unable to investigate the friendliness between different types of control algorithms.

\subsection{Real network results}
\begin{figure}
\begin{centering}
\includegraphics[width=0.45\textwidth]{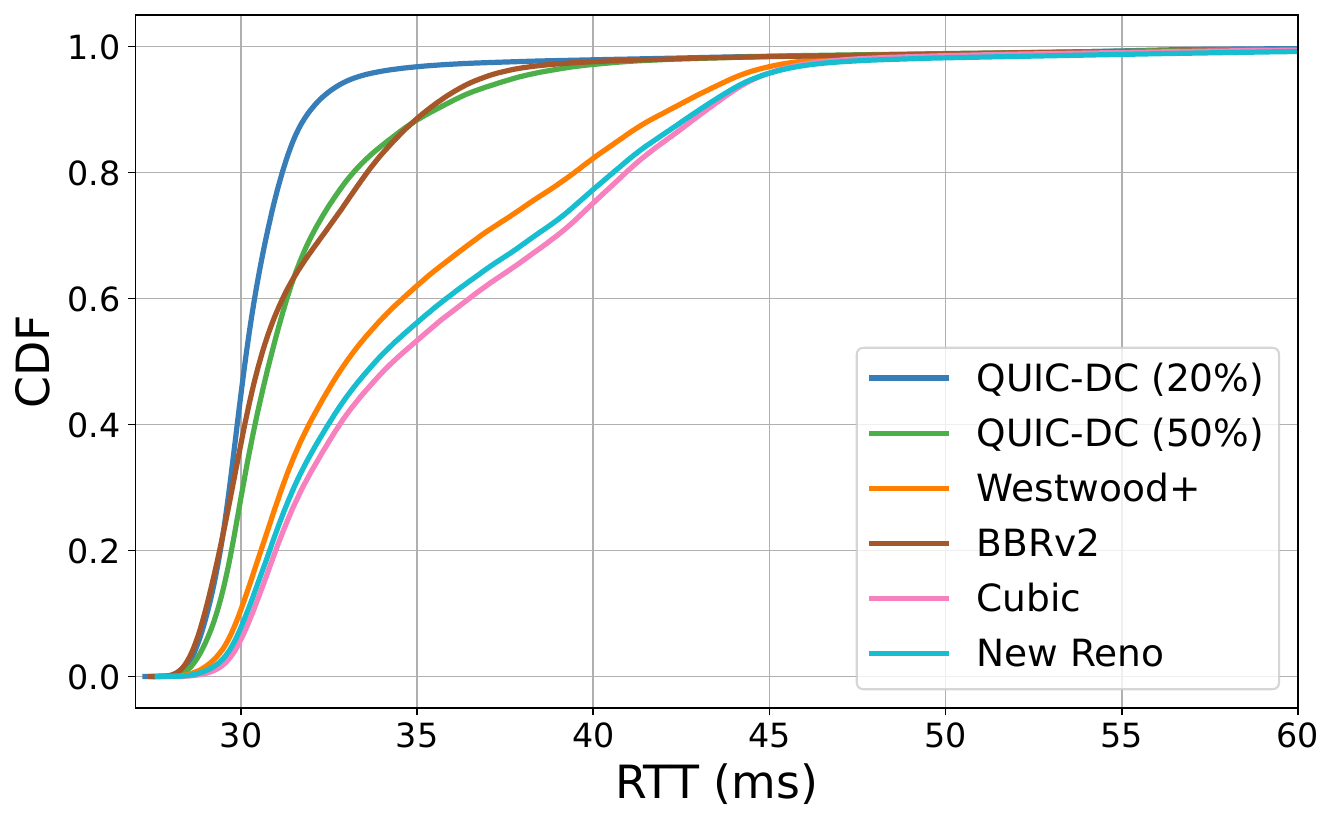}
\par\end{centering}
\vspace{-3mm}
\caption{RTT distribution on real network path. }
\label{fig:real-net-single-flow}
\vspace{-3mm}
\end{figure}

We consider the real network scenario described in Section~\ref{sec:testbed}. During the experiments, we have measured $RTT{min}=30 \ ms$, $RTT{max}=60 \ ms$, which means that there is a bottleneck buffer size equal to $1 \ BDP$. Table \ref{tab:realtable} shows that
QUIC-DC reduces the loss rate by one order of magnitude whereas the goodput is only slightly lower than the other approaches (note that QUIC Westwood+ corresponds to QUIC-DC $(100\%)$). Moreover, Figure \ref{fig:real-net-single-flow} shows that QUIC-DC also provides shorter RTT. However,  it should also be noted that algorithms experiencing higher loss rates provide RTT delays that are underestimated because retransmitted packets are not considered in RTT measurements~\cite{karn1987improving}. In other terms, the algorithms with larger loss-rates provide a distribution of RTT that underestimates the effective delay.

\begin{table}
\centering
\caption{Single flow, real network connections.}
\label{tab:realtable}
\resizebox{0.48\textwidth}{!}{
\begin{tabular}{lrrrrr}
\toprule
\textbf{CCA} & \textbf{Gput} & \textbf{Tput} & \textbf{Loss} & \textbf{RTT$_{avg}$} & \textbf{RTT$_{std}$} \\
            & (Mbps)        & (Mbps)        & (\%)          & (ms)                 & (ms) \\
\midrule
QUIC-DC (20\%)        & 26.744 & 26.746 & 0.0096 & 30.92 & 9.30  \\
QUIC-DC (50\%)        & 30.954 & 30.956 & 0.0103 & 32.09 & 11.18 \\
Westwood+              & 33.115 & 33.132 & 0.0470 & 35.08 & 11.50 \\
BBRv2                 & 33.560 & 33.586 & 0.0730 & 32.00 & 11.28 \\
Cubic                 & 33.529 & 33.555 & 0.0733 & 36.17 & 12.05 \\
New Reno              & 33.935 & 32.959 & 0.0765 & 35.92 & 12.18 \\
\bottomrule
\end{tabular}
}
\end{table}


\section{Conclusions}
QUIC DC  controls the delay of an Internet connection by measuring the one-way queueing delay to trigger an early reaction to congestion. The idea has been implemented in QUIC using the TCP Westwood+ congestion control and has been evaluated against QUIC NewReno, QUIC Westwood+, QUIC Cubic, and QUIC BBRv2. Experimental results show that the cumulative distribution function of the round-trip time can be shifted to lower values, while the packet loss rate can be reduced by an order of magnitude. BBRv2 has shown serious drawbacks when multiple flows share the bottleneck.

\begin{acks}
This work has been funded by the “LOREN” Project,\\ n.\ 20223Y85JN, under the PRIN (Progetti di Ricerca d’Interesse Nazionale) 2022 programme of the Italian Ministry of Research.
\end{acks}

\bibliographystyle{ACM-Reference-Format}
\bibliography{biblio}


\begin{thebibliography}{21}


\ifx \showCODEN    \undefined \def \showCODEN     #1{\unskip}     \fi
\ifx \showISBNx    \undefined \def \showISBNx     #1{\unskip}     \fi
\ifx \showISBNxiii \undefined \def \showISBNxiii  #1{\unskip}     \fi
\ifx \showISSN     \undefined \def \showISSN      #1{\unskip}     \fi
\ifx \showLCCN     \undefined \def \showLCCN      #1{\unskip}     \fi
\ifx \shownote     \undefined \def \shownote      #1{#1}          \fi
\ifx \showarticletitle \undefined \def \showarticletitle #1{#1}   \fi
\ifx \showURL      \undefined \def \showURL       {\relax}        \fi
\providecommand\bibfield[2]{#2}
\providecommand\bibinfo[2]{#2}
\providecommand\natexlab[1]{#1}
\providecommand\showeprint[2][]{arXiv:#2}

\bibitem[Brakmo and Peterson(1995)]%
        {vegas}
\bibfield{author}{\bibinfo{person}{L.S. Brakmo} {and} \bibinfo{person}{L.L. Peterson}.} \bibinfo{year}{1995}\natexlab{}.
\newblock \showarticletitle{TCP Vegas: end to end congestion avoidance on a global Internet}.
\newblock \bibinfo{journal}{\emph{IEEE Journal on Selected Areas in Communications}} \bibinfo{volume}{13}, \bibinfo{number}{8} (\bibinfo{year}{1995}), \bibinfo{pages}{1465--1480}.
\newblock
\href{https://doi.org/10.1109/49.464716}{doi:\nolinkurl{10.1109/49.464716}}


\bibitem[Budzisz et~al\mbox{.}(2011)]%
        {bun11}
\bibfield{author}{\bibinfo{person}{Łukasz Budzisz}, \bibinfo{person}{Rade Stanojevic}, \bibinfo{person}{Arieh Schlote}, \bibinfo{person}{Fred Baker}, {and} \bibinfo{person}{Robert Shorten}.} \bibinfo{year}{2011}\natexlab{}.
\newblock \showarticletitle{On the Fair Coexistence of Loss- and Delay-Based TCP}.
\newblock \bibinfo{journal}{\emph{IEEE/ACM Transactions on Networking}} \bibinfo{volume}{19}, \bibinfo{number}{6} (\bibinfo{year}{2011}), \bibinfo{pages}{1811--1824}.
\newblock
\href{https://doi.org/10.1109/TNET.2011.2159736}{doi:\nolinkurl{10.1109/TNET.2011.2159736}}


\bibitem[Cardwell et~al\mbox{.}(2019)]%
        {cardwell2019bbr}
\bibfield{author}{\bibinfo{person}{Neal Cardwell}, \bibinfo{person}{Yuchung Cheng}, \bibinfo{person}{Soheil~Hassas Yeganeh}, \bibinfo{person}{Priyaranjan Jha}, \bibinfo{person}{Yousuk Seung}, \bibinfo{person}{Ian Swett}, \bibinfo{person}{Victor Vasiliev}, \bibinfo{person}{Bin Wu}, \bibinfo{person}{Matt Mathis}, {and} \bibinfo{person}{Van Jacobson}.} \bibinfo{year}{2019}\natexlab{}.
\newblock \showarticletitle{BBR v2: a model-based congestion control IETF 105 update}.
\newblock \bibinfo{journal}{\emph{Presentation at IETF105}} (\bibinfo{year}{2019}).
\newblock


\bibitem[Carlucci et~al\mbox{.}(2017)]%
        {gcc}
\bibfield{author}{\bibinfo{person}{Gaetano Carlucci}, \bibinfo{person}{Luca De~Cicco}, \bibinfo{person}{Stefan Holmer}, {and} \bibinfo{person}{Saverio Mascolo}.} \bibinfo{year}{2017}\natexlab{}.
\newblock \showarticletitle{Congestion control for web real-time communication}.
\newblock \bibinfo{journal}{\emph{IEEE/ACM Transactions on Networking}} \bibinfo{volume}{25}, \bibinfo{number}{5} (\bibinfo{year}{2017}), \bibinfo{pages}{2629--2642}.
\newblock


\bibitem[Carofiglio et~al\mbox{.}(2010)]%
        {car10}
\bibfield{author}{\bibinfo{person}{Giovanna Carofiglio}, \bibinfo{person}{Luca Muscariello}, \bibinfo{person}{Dario Rossi}, {and} \bibinfo{person}{Silvio Valenti}.} \bibinfo{year}{2010}\natexlab{}.
\newblock \showarticletitle{The quest for LEDBAT fairness}. In \bibinfo{booktitle}{\emph{Proc. IEEE GLOBECOM '10}}. IEEE.
\newblock


\bibitem[Grieco and Mascolo(2004)]%
        {gri04}
\bibfield{author}{\bibinfo{person}{Luigi~A. Grieco} {and} \bibinfo{person}{Saverio Mascolo}.} \bibinfo{year}{2004}\natexlab{}.
\newblock \showarticletitle{Performance evaluation and comparison of Westwood+, New Reno, and Vegas TCP congestion control}.
\newblock \bibinfo{journal}{\emph{SIGCOMM Comput. Commun. Rev.}} \bibinfo{volume}{34}, \bibinfo{number}{2} (\bibinfo{date}{April} \bibinfo{year}{2004}), \bibinfo{pages}{25–38}.
\newblock
\showISSN{0146-4833}
\href{https://doi.org/10.1145/997150.997155}{doi:\nolinkurl{10.1145/997150.997155}}


\bibitem[Ha et~al\mbox{.}(2008)]%
        {ha2008cubic}
\bibfield{author}{\bibinfo{person}{Sangtae Ha}, \bibinfo{person}{Injong Rhee}, {and} \bibinfo{person}{Lisong Xu}.} \bibinfo{year}{2008}\natexlab{}.
\newblock \showarticletitle{CUBIC: a new TCP-friendly high-speed TCP variant}.
\newblock \bibinfo{journal}{\emph{ACM SIGOPS operating systems review}} \bibinfo{volume}{42}, \bibinfo{number}{5} (\bibinfo{year}{2008}), \bibinfo{pages}{64--74}.
\newblock


\bibitem[Hayes and Armitage(2011)]%
        {hay11}
\bibfield{author}{\bibinfo{person}{David~A Hayes} {and} \bibinfo{person}{Grenville Armitage}.} \bibinfo{year}{2011}\natexlab{}.
\newblock \showarticletitle{Revisiting TCP congestion control using delay gradients}. In \bibinfo{booktitle}{\emph{International Conference on Research in Networking}}. Springer, \bibinfo{pages}{328--341}.
\newblock


\bibitem[Henderson et~al\mbox{.}(2012)]%
        {henderson2012newreno}
\bibfield{author}{\bibinfo{person}{Tom Henderson}, \bibinfo{person}{Sally Floyd}, \bibinfo{person}{Andrei Gurtov}, {and} \bibinfo{person}{Yoshifumi Nishida}.} \bibinfo{year}{2012}\natexlab{}.
\newblock \bibinfo{booktitle}{\emph{The NewReno modification to TCP's fast recovery algorithm}}.
\newblock \bibinfo{type}{{T}echnical {R}eport}.
\newblock


\bibitem[Iyengar and Thomson(2021)]%
        {rfc9000}
\bibfield{author}{\bibinfo{person}{J. Iyengar} {and} \bibinfo{person}{M. Thomson}.} \bibinfo{year}{2021}\natexlab{}.
\newblock \bibinfo{booktitle}{\emph{QUIC: A UDP-Based Multiplexed and Secure Transport}}.
\newblock \bibinfo{type}{RFC} 9000. \bibinfo{institution}{RFC Editor}.
\newblock
\urldef\tempurl%
\url{https://www.rfc-editor.org/rfc/rfc9000.txt}
\showURL{%
\tempurl}


\bibitem[Jacobson(1988)]%
        {jac88}
\bibfield{author}{\bibinfo{person}{V. Jacobson}.} \bibinfo{year}{1988}\natexlab{}.
\newblock \showarticletitle{Congestion avoidance and control}. In \bibinfo{booktitle}{\emph{Proc. of ACM SIGCOMM '88}} (Stanford, California, USA) \emph{(\bibinfo{series}{SIGCOMM '88})}. \bibinfo{publisher}{Association for Computing Machinery}, \bibinfo{address}{New York, NY, USA}, \bibinfo{pages}{314–329}.
\newblock
\showISBNx{0897912799}
\href{https://doi.org/10.1145/52324.52356}{doi:\nolinkurl{10.1145/52324.52356}}


\bibitem[Jain(1989)]%
        {jain89}
\bibfield{author}{\bibinfo{person}{R. Jain}.} \bibinfo{year}{1989}\natexlab{}.
\newblock \showarticletitle{A delay-based approach for congestion avoidance in interconnected heterogeneous computer networks}.
\newblock \bibinfo{journal}{\emph{SIGCOMM Comput. Commun. Rev.}} \bibinfo{volume}{19}, \bibinfo{number}{5} (\bibinfo{date}{Oct.} \bibinfo{year}{1989}), \bibinfo{pages}{56–71}.
\newblock
\showISSN{0146-4833}
\href{https://doi.org/10.1145/74681.74686}{doi:\nolinkurl{10.1145/74681.74686}}


\bibitem[Karn and Partridge(1987)]%
        {karn1987improving}
\bibfield{author}{\bibinfo{person}{Phil Karn} {and} \bibinfo{person}{Craig Partridge}.} \bibinfo{year}{1987}\natexlab{}.
\newblock \showarticletitle{Improving round-trip time estimates in reliable transport protocols}.
\newblock \bibinfo{journal}{\emph{ACM SIGCOMM Computer Communication Review}} \bibinfo{volume}{17}, \bibinfo{number}{5} (\bibinfo{year}{1987}), \bibinfo{pages}{2--7}.
\newblock


\bibitem[Mascolo et~al\mbox{.}(2001)]%
        {mascolo2001tcp}
\bibfield{author}{\bibinfo{person}{Saverio Mascolo}, \bibinfo{person}{Claudio Casetti}, \bibinfo{person}{Mario Gerla}, \bibinfo{person}{Medy~Y Sanadidi}, {and} \bibinfo{person}{Ren Wang}.} \bibinfo{year}{2001}\natexlab{}.
\newblock \showarticletitle{TCP Westwood: Bandwidth estimation for enhanced transport over wireless links}. In \bibinfo{booktitle}{\emph{Proceedings of the 7th annual international conference on Mobile computing and networking}}. \bibinfo{pages}{287--297}.
\newblock


\bibitem[Parsa and Garcia-Luna-Aceves(1999)]%
        {parsa1999improving}
\bibfield{author}{\bibinfo{person}{Christina Parsa} {and} \bibinfo{person}{Jose~Joaquin Garcia-Luna-Aceves}.} \bibinfo{year}{1999}\natexlab{}.
\newblock \showarticletitle{Improving TCP congestion control over internets with heterogeneous transmission media}. In \bibinfo{booktitle}{\emph{Proc. ICNP '99}}. IEEE, \bibinfo{pages}{213--221}.
\newblock


\bibitem[Peterson and Davie(2021)]%
        {peterson2007computer}
\bibfield{author}{\bibinfo{person}{Larry~L Peterson} {and} \bibinfo{person}{Bruce~S Davie}.} \bibinfo{year}{2021}\natexlab{}.
\newblock \bibinfo{booktitle}{\emph{Computer networks: a systems approach}}.
\newblock \bibinfo{publisher}{Elsevier}.
\newblock


\bibitem[Prasad et~al\mbox{.}(2004)]%
        {dr04}
\bibfield{author}{\bibinfo{person}{Ravi~S Prasad}, \bibinfo{person}{Manish Jain}, {and} \bibinfo{person}{Constantinos Dovrolis}.} \bibinfo{year}{2004}\natexlab{}.
\newblock \showarticletitle{On the effectiveness of delay-based congestion avoidance}. In \bibinfo{booktitle}{\emph{Proc. PFLDNet}}, Vol.~\bibinfo{volume}{4}.
\newblock


\bibitem[Ramakrishnan et~al\mbox{.}(2001)]%
        {ramakrishnan2001addition}
\bibfield{author}{\bibinfo{person}{Kadangode Ramakrishnan}, \bibinfo{person}{Sally Floyd}, {and} \bibinfo{person}{David Black}.} \bibinfo{year}{2001}\natexlab{}.
\newblock \bibinfo{booktitle}{\emph{The addition of explicit congestion notification (ECN) to IP}}.
\newblock \bibinfo{type}{{T}echnical {R}eport}.
\newblock


\bibitem[Shalunov et~al\mbox{.}(2012)]%
        {sha12}
\bibfield{author}{\bibinfo{person}{Sea Shalunov}, \bibinfo{person}{Greg Hazel}, \bibinfo{person}{Janardhan Iyengar}, {and} \bibinfo{person}{Mirja Kuehlewind}.} \bibinfo{year}{2012}\natexlab{}.
\newblock \bibinfo{booktitle}{\emph{Low extra delay background transport (LEDBAT)}}.
\newblock \bibinfo{type}{RFC}. \bibinfo{institution}{RFC Editor}.
\newblock
\urldef\tempurl%
\url{https://www.rfc-editor.org/rfc/rfc6817.txt}
\showURL{%
\tempurl}


\bibitem[Zaki et~al\mbox{.}(2015)]%
        {zak15}
\bibfield{author}{\bibinfo{person}{Yasir Zaki}, \bibinfo{person}{Thomas P{\"o}tsch}, \bibinfo{person}{Jay Chen}, \bibinfo{person}{Lakshminarayanan Subramanian}, {and} \bibinfo{person}{Carmelita G{\"o}rg}.} \bibinfo{year}{2015}\natexlab{}.
\newblock \showarticletitle{Adaptive congestion control for unpredictable cellular networks}. In \bibinfo{booktitle}{\emph{Proc. ACM SIGCOMM '15}}. \bibinfo{pages}{509--522}.
\newblock


\bibitem[Zeynali et~al\mbox{.}(2024)]%
        {zey24}
\bibfield{author}{\bibinfo{person}{Danesh Zeynali}, \bibinfo{person}{Emilia~N. Weyulu}, \bibinfo{person}{Seifeddine Fathalli}, \bibinfo{person}{Balakrishnan Chandrasekaran}, {and} \bibinfo{person}{Anja Feldmann}.} \bibinfo{year}{2024}\natexlab{}.
\newblock \showarticletitle{Promises and Potential of BBRv3}. In \bibinfo{booktitle}{\emph{Proc. of PAM 2024}}. \bibinfo{publisher}{Springer-Verlag}, \bibinfo{address}{Berlin, Heidelberg}, \bibinfo{pages}{249–272}.
\newblock
\showISBNx{978-3-031-56251-8}
\href{https://doi.org/10.1007/978-3-031-56252-5_12}{doi:\nolinkurl{10.1007/978-3-031-56252-5_12}}


\end{thebibliography}

\end{document}